\documentclass{PoS}

\usepackage{graphicx}
\usepackage{fixmath}
\pdfoutput=1

\title{Summary of WG6: Spin Physics}

\ShortTitle{Summary of WG6: Spin Physics}

\author{Emanuele R. Nocera\\
        Rudolf Peierls Centre for Theoretical Physics\\
        1 Keble Road, University of Oxford\\
        OX1 3NP Oxford, United Kingdom\\
        E-mail: \email{emanuele.nocera@physics.ox.ac.uk}}
\author{Silvia Pisano\\
        INFN Laboratori Nazionali di Frascati\\ 
        Via Enrico Fermi 40, 00044, Frascati, Italy\\
        E-mail: \email{silvia.pisano@lnf.infn.it}}

\abstract{The working group on Spin Physics at the XXIV International 
Workshop on Deep-Inelastic Scattering and Related Subjects (DIS2016) 
witnessed a significant progress in the theoretical and experimental 
investigations aiming at unveiling the innermost structure of the proton. 
Results ranged from proton's one-dimensional representation to its 
multi-dimensional imaging. In this contribution, we summarize a selection 
of the topics discussed and of the results presented. For details, 
we refer to the individual contributions collected in the proceedings of 
this workshop.}

\FullConference{XXIV International Workshop on Deep-Inelastic Scattering and Related Subjects\\
		11-15 April, 2016\\
		DESY Hamburg, Germany}

\begin{document}

\section{Introduction}
\label{sec:introduction}

Nucleons, protons and neutrons, are those bound states which make up all nuclei,
and hence most of the visible matter in the Universe. Understanding their 
fundamental structure and dynamics in terms of their partonic constituents - 
quarks and gluons - is currently one of the main challenges in hadron 
physics~\cite{Thomas:2001kw}. Such an understanding has been rooted
in the theoretical framework provided by Quantum Chromodynamics (QCD), the 
field theory which describes the strong interaction among 
quarks and gluons~\cite{Gross:1973ju}. The theory has been 
developed to match the experimental probe of nucleon structure, which consists 
in scattering nucleons off a beam of leptons, or of other nucleons, in a 
large-momentum transfer process. Following factorization~\cite{Collins:1989gx}, 
hadronic cross sections measured in sufficiently inclusive processes are 
described as a convolution between a short-distance part - that encodes 
information on the hard interactions of partons in the form of perturbative 
computable, process-dependent kernels - and a long-distance part - that 
encodes information on the longitudinal momentum structure of the nucleon in 
the form of universal parton distribution functions (PDFs). 

Despite this framework has been tremendously successful in the quantitative
description of a wealth of experimental data measured by a number of facilities 
around the world, some fundamental aspects of the partonic structure of the
nucleon are still rather poorly determined. The focus of hadron physics is 
currently on two such aspects: on the one hand, the understanding of the 
innermost nature of nucleon's total angular momentum in terms of the 
individual contributions of quarks, antiquarks and gluons~\cite{Aidala:2012mv}; 
on the other hand, the quest of a multi-dimensional picture of the nucleon, in 
which parton's transverse momentum and spatial position are taken into 
account~\cite{Anselmino:2016mv}.

\begin{figure}[t]
\centering
\includegraphics[width=\textwidth]{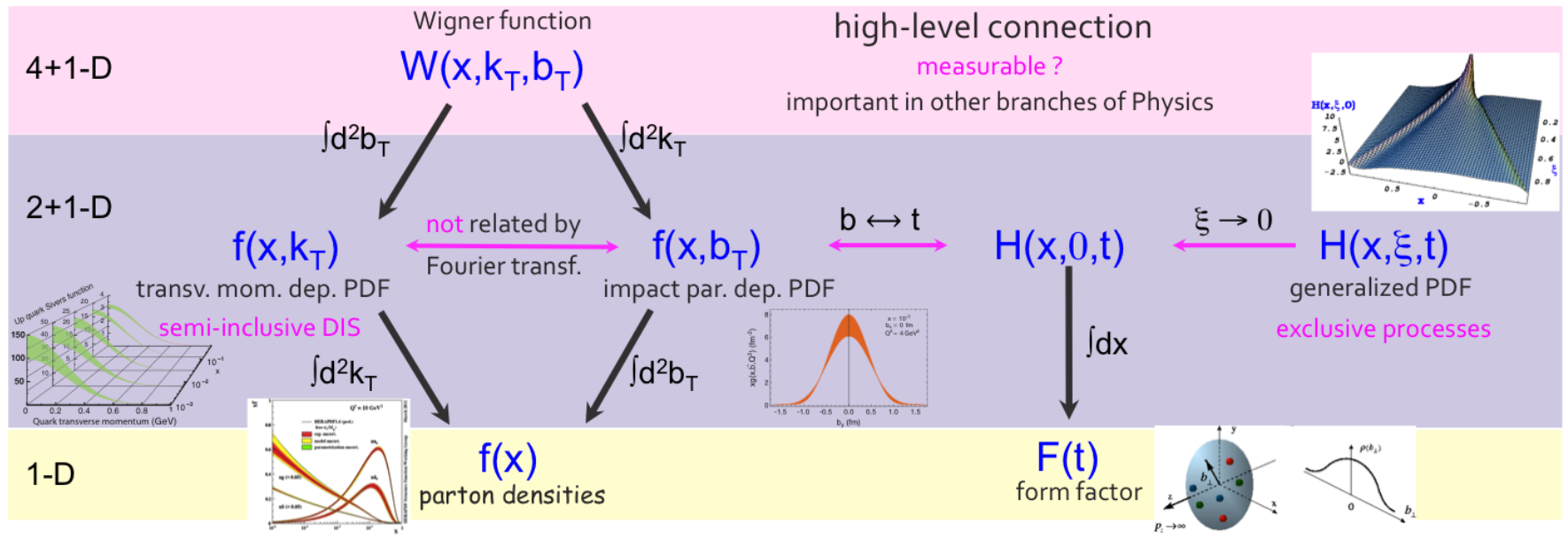}
\caption{Connections among different functions describing the distribution of
partons inside the proton. The functions given here are for unpolarized 
partons in an unpolarized proton; analogous relations hold for polarized
quantities. The figure is taken from Ref.~\cite{Aschenauer:2015eha}.}
\label{fig:PDFs}
\end{figure}

These two aspects are strictly entwined with each other, as it has become clear
for instance in the decomposition of the nucleon total angular momentum,
of which a possible realization~\cite{Jaffe:1989jz}
in terms of its quark and gluon spin, $\Delta\Sigma$ 
and $\Delta G$, and quark and gluon orbital momentum, $\mathcal{L}_q$ and 
$\mathcal{L}_g$, is
\begin{equation}
\frac{1}{2}
=
\frac{1}{2}\Delta\Sigma (\mu^2) + \Delta G (\mu^2) 
+
\mathcal{L}_q(\mu^2) + \mathcal{L}_g(\mu^2)
\, ,
\label{eq:spinsumrule}
\end{equation}
where $\mu^2$ is the factorization scale. In principle one would like to be able
to measure each term in Eq.~(\ref{eq:spinsumrule}), or at least a combination
of them.\footnote{Note that the decomposition provided by 
Eq.~(\ref{eq:spinsumrule}) is not unique. What should be the decompositions 
that lead to gauge-invariant, physically meaningful terms 
(and in which sense these
are measurable) are discussed in Ref.~\cite{Leader:2013jra}.}
While $\Delta\Sigma$ and $\Delta G$, which encode the information on the 
one-dimensional (1-D) spin structure of the proton, can be related to the 
longitudinally polarized PDFs, $\mathcal{L}_q$ and $\mathcal{L}_g$, which 
encode the information on the multi-dimensional structure of the proton, 
should be related to some different objects. In principle, the distribution of 
partons inside a nucleon, as a simultaneous function of their longitudinal 
momentum fraction $x$, transverse momentum $\mathbold{k_T}$ and 
impact-parameter space $\mathbold{b_T}$, is encoded in the five-dimensional 
(5-D) Wigner distribution $W(x,\mathbold{k_T},\mathbold{b_T})$. However, this 
cannot be accessed in the experiment. Integrating 
$W(x,\mathbold{k_T},\mathbold{b_T})$ in $\mathbold{b_T}$ leads to the 
three-dimensional (3-D) transverse momentum dependent
PDFs (TMDs), which are instead measurable quantities (typically in 
semi-inclusive processes). Note that a further integration of the TMDs in 
$\mathbold{k_T}$ leads to the usual 1-D PDFs. Otherwise, integrating 
$W(x,\mathbold{k_T},\mathbold{b_T})$ in 
$\mathbold{k_T}$, then Fourier transforming the $\mathbold{b_T}$ dependence 
into the Mandelstam variable $-t$, and finally 
extrapolating from zero to finite skewness $\xi$ leads to the 3-D generalized 
parton distribution functions (GPDs), see Fig.~\ref{fig:PDFs}. These are 
measurable objects too (typically in exclusive processes), and may be related 
to the quark and gluon orbital momentum~\cite{Boffi:2007yc}.

The working group on Spin Physics (WG6) at the XXIV International Workshop
on Deep-Inelastic Scattering and Related Subjects (DIS2016) 
addressed recent theoretical and experimental developments in hadron 
physics, including the subjects outlined above. Almost fourteen hours, 
divided into eight sessions, were dedicated to presentations and discussions. 
Invited overview talks were also called for.
An additional two-hour session, joint with the working group on 
Future Experiments (WG7), was organized to focus on prospective developments 
in spin physics, connected to future experimental facilities. 
A total of thirty-eight talks were presented.

In this contribution, we summarize some selected results: in Sec.~\ref{sec:1D}, 
we discuss the 1-D representation of the longitudinal spin structure of the 
nucleon, while in Sec.~\ref{sec:3D} we address the 3-D mapping of the nucleon, 
focusing on TMDs and GPDs. Further details on each topic can be found in the 
individual contributions in these proceedings.

\section{1-D representation of the longitudinal nucleon spin structure}
\label{sec:1D}

The first two terms in Eq.~(\ref{eq:spinsumrule}) represent the fraction of
the nucleon spin carried by quarks and gluons respectively. Provided that
the longitudinally polarized PDFs of the proton
\begin{equation}
\Delta f(x,\mu^2) \equiv f^\uparrow(x,\mu^2)  - f^\downarrow(x,\mu^2)\, 
\label{eq:polPDFs}
\end{equation}
are defined as the net momentum densities of partons ($f$ denotes either a 
quark $q$, an antiquark $\bar{q}$, or a gluon $g$) with spin aligned along 
($\uparrow$) or opposite ($\downarrow$) the polarization of the parent nucleon, 
$\Delta\Sigma$ and $\Delta G$ are then defined as their first moments
\begin{equation}
\Delta\Sigma(\mu^2) 
= 
\sum_{q=u,d,s} \int_0^1 dx\,\left[\Delta q(x,\mu^2) + \Delta\bar{q}(x,\mu^2)\right]\, ,
\ \ \ \ \
\Delta G(x,\mu^2)
=
\int_0^1 dx\, \Delta g(x,\mu^2)\, .
\label{eq:fmom}
\end{equation} 
The dependence of the PDFs on the factorization scale $\mu$ can be obtained
as a perturbative solution of the DGLAP evolution 
equations~\cite{Altarelli:1977zs}, and the coefficients of the corresponding
splitting functions have been computed up to next-to-next-to-leading order
(NNLO) accuracy very recently~\cite{Moch:2014sna}. The dependence of the PDF
on the proton momentum fraction $x$ carried by the parton is nonperturbative,
and must be determined from the data. This is supplemented with some
theoretical requirements, which include: positivity, {\it i.e.} PDFs
must lead to positive cross sections; integrability, {\it i.e.} the 
nucleon matrix element of the axial current must be finite for each flavor;
SU(2) and SU(3) flavor symmetry, {\it i.e.} the first moments of the 
nonsinglet $\mathcal{C}$-even PDF combinations are related to the baryon
octed $\beta$-decay constants, whose values are well 
measured~\cite{Agashe:2014kda}. Recent theoretical and experimental progress in 
the determinations of the PDFs was extensively reported in WG6.

On the theoretical side, the available sets of longitudinally polarized PDFs
of the proton differ among each other for the procedure used to determine PDFs 
from the data, for the data set included, and for the details of the QCD analysis,
see {\it e.g.} Chap.~3 in Ref.~\cite{Nocera:2014vla} for a review. 

As far as the procedure is concerned, three different approaches are being 
used. The first is based on simple polynomials for PDF parametrization and 
on standard Hessian or Lagrange multiplier techniques for PDF uncertainty 
estimates. The second is based on neural networks for PDF parametrization 
and Monte Carlo sampling for PDF uncertainty estimates. The third combines 
the two approaches, keeping simple polynomials for PDF parametrization and 
Monte Carlo sampling for PDF uncertainty estimates. Monte Carlo sampling
techniques have the advantage to generate fits of PDFs with statistically
rigorous uncertainties, while neural networks allow for a reduction of 
the bias associated to the choice of the parametrization.
Among recent fits, those from the {\tt DSSV} 
family~\cite{deFlorian:2008mr,deFlorian:2009vb,deFlorian:2014yva} are based 
on the first approach, those from the {\tt NNPDF} 
collaboration~\cite{Ball:2013lla,Nocera:2014gqa}
are based on the second and that from the {\tt JAM} 
collaboration~\cite{Sato:2016tuz} is based on the third.

As far as the data set is concerned, the bulk of the experimental information 
in all the above-mentioned fits is provided by neutral current, photon-induced, 
inclusive deep-inelastic scattering (DIS). Very recent DIS measurements
from COMPASS and JLAB have been included in a dedicated {\tt NNPDF} 
analysis~\cite{Nocera:2015vva} as well as in the {\tt JAM} PDF set. 
Because of the way the corresponding observables factorize, inclusive DIS data 
only constrain the total quark-antiquark combinations and, weakly, the gluon
distribution via scaling violations. In order to overcome this issue, 
observables sensitive to individual quark and antiquark polarizations
and receiving a leading contribution from gluon-initiated partonic subprocesses
should be considered respectively. In the 
{\tt DSSV} analysis the quark-antiquark PDFs are determined from 
semi-inclusive DIS (SIDIS), while the gluon PDF is determined from 
pion and jet production in polarized proton-proton ($pp$) collision at the 
Relativistic Heavy Ion Collider (RHIC); in the {\tt NNPDF} analysis the
quark-antiquark PDFs are determined from $W$ boson production at RHIC, 
while the gluon PDF is determined from open-charm leptoproduction in SIDIS
and jet production at RHIC. The inclusion of 
SIDIS and pion production data requires the usage of poorly known 
fragmentation functions (FFs), which may spoil the accuracy of the ensued PDFs.
Overall, RHIC data were demonstrated to point towards a sizable, positive 
gluon polarization in the proton~\cite{deFlorian:2014yva,Nocera:2014gqa}, 
and a sizable asymmetry of the polarized sea quarks~\cite{Nocera:2014rea}, 
though these conclusions hold only in the rather limited region covered by data,
$0.05\lesssim x \lesssim 0.5$.

As far as the details of the QCD analysis are concerned, the {\tt DSSV},
{\tt NNPDF} and {\tt JAM} PDF sets are determined in the $\overline{\rm MS}$ 
scheme at next-to-leading order (NLO) accuracy.\footnote{A NNLO QCD 
analysis of longitudinally polarized PDFs has been performed very 
recently, see Ref.~\cite{Shahri:2016uzl}.} In the {\tt JAM} analysis,
particular attention was put in the inclusion of finite-$\mu^2$ 
and nuclear corrections relevant for JLAB kinematics; flavor-separated
twist-3 contributions and the $d_2$ moment of the nucleon were also determined.
Although all the available PDF sets are determined at fixed order in the 
perturbative QCD expansion, resummed PDF fits are in priniple 
possible~\cite{Ringer:2015pd}, and threshold resummation for open-charm
production at COMPASS was specifically discussed~\cite{Uebler:2015ria}.
A valuable input on the determination of PDFs may come from nonperturbative 
models of nucleon structure~\cite{Nocera:2014uea}, and, more and more
promisingly, from first principle computations in lattice 
QCD~\cite{Wiese:2016dis}.

On the experimental side, a wealth of new experimental data was illustrated. 

The COMPASS collaboration presented new measurements of the deuteron 
longitudinal double spin asymmetry $A_1^d$ and of the deuteron spin-dependent 
structure function $g_1^d$ in DIS~\cite{Wilfert:2016dis}. These data sets 
provide twice the precision of earlier COMPASS measurements and complement 
the analogous results for protons~\cite{Adolph:2015saz}. The COMPASS 
collaboration also presented a new analysis of the proton longitudinal spin 
asymmetry, $A_1^p$, and structure function, $g_1^p$, at low values of $x$ and 
$\mu^2$ in two-dimensional bins of various kinematic 
variables~\cite{Nunes:2016dis}. These data sets point towards a first evidence 
of positive spin effects at low $x$, which seems to be independent of the 
kinematics. 

The PHENIX collaboration presented a new measurement of the double spin 
asymmetry for mid-rapidity neutral pion production in polarized $pp$ 
collisions at a center-of-mass energy $\sqrt{s}=510$ GeV~\cite{Adare:2015ozj}. 
This data set confirms a positive contribution of the gluon polarization to 
the proton spin, consistent with the finding from earlier measurements of 
pion and jet asymmetries; it also extends the sensitivity to $\Delta G$ down to
$x\sim 0.01$ into a currently unexplored region. The PHENIX collaboration
also reported on the double helicity asymmetry for inclusive $J/\Psi$ 
production at forward rapidity in $\sqrt{s}=510$ GeV polarized $pp$ 
collisions~\cite{Adare:2016cqe}. Because $J/\Psi$ particles are predominantly 
produced through gluon-gluon scattering, this data set is a further handle on 
$\Delta G$, especially in the low $x$ region, provided that the mechanism
of $J/\Psi$ production is clarified. The PHENIX collaboration also presented
forward- and mid-rapidity measurements for $W^\pm/Z$ production in 
polarized $pp$ collisions~\cite{Adare:2015gsd}. 
This data set will provide further insight into antiquark 
polarized PDFs.

The STAR collaboration presented first dijet asymmetry measurements in 
polarized $pp$ collisions at $\sqrt{s}=500$ GeV and $\sqrt{s}=510$ 
GeV~\cite{Ramachandran:2016dis}. This data set complements earlier STAR
measurements on single-jet production, and will extend the current constraints 
on $\Delta G$ down to $x\sim 0.01$.

Despite all these achievements, the lack of 
experimental information over a wide range of $x$ and $\mu^2$ seriously 
limits the accuracy with which $\Delta\Sigma$ and $\Delta G$ in 
Eq.~(\ref{eq:spinsumrule}) can be determined~\cite{Nocera:2015fxa}. 
Brand new facilities, such as the proposed high-energy, polarized Electron-Ion 
Collider (EIC)~\cite{Accardi:2012qut} will be required to definetely unveil 
their size with few percent accuracy~\cite{Aschenauer:2015ata,Ball:2013tyh}.

\section{3-D mapping of the nucleon}
\label{sec:3D}

Once the 1-D representation of the nucleon structure is extended to a full 3-D imaging, 
a plethora of new phenomena emerge, connected to the non-trivial parton dynamics in the hadrons.

A first 3-D representation of the nucleon follows from the combination of the electromagnetic 
form factor and the PDF. The former describes the transverse distribution of the nucleon charge 
and magnetization, while the latter encodes the information on how the nucleon momentum is shared
among its constituents 
in the longitudinal direction. The combination of the two leads to the GPDs, which describe 
the correlation among the parton transverse position and their momentum fraction (see 
Fig.~\ref{fig:PDFs}). 
At leading twist, four different GPDs describe the nucleon for each quark flavor (in the \textit{chiral-even} sector), 
depending on the occurrence or not of a spin flip in the quark after the interaction with the hard 
probe~\cite{Boffi:2007yc}. They are usually accessed through deeply-virtual Compton scattering
(DVCS), a hard exclusive process
consisting in the electro-production of a \textit{real} photon off a nucleon target. 
From a theoretical point of view, DVCS is described through the so-called \textit{handbag} diagram, 
where the hard, perturbative dynamics describing the interaction of the struck quark with the
exchanged 
virtual photon is factorized out of the soft, long-range dynamics describing the hadronic bound
state. 
Such a soft, nonperturbative part is encoded in the GPDs: they depend on the fraction $x$ of the 
nucleon momentum carried by the interacting quark, on the four-momentum 
$-t$ exchanged between the electron and the proton, and on its tranverse fraction $\xi$.

The analysis of the DVCS cross-section $-t$-slope provides a mapping of the nucleon size as a 
function of the fraction $x$ considered, leading to an actual \textit{transverse nucleon imaging}. 
While high-$x$, valence quarks seem to be confined at the center of the nucleon, low-$x$ contributions
seem to be 
spread at the periphery, resulting in a larger nucleon size once the low-$x$ region is explored. 
A recent measurement by COMPASS on the 2012 pilot run has been reported, which extends the previous 
measurements 
by H1 and ZEUS~\cite{Chekanov:2008vy,Aktas:2005ty,Aaron:2009ac} in the low-$x$ region. 
Furthermore, the latest JLab data from Hall-A \cite{Camacho:2006qlk,Defurne:2015kxq} 
and Hall-B \cite{Jo:2015ema} cover with unprecedented precision the high-$x$ domain~\cite{Niccolai:2016dis}. 
Finally, a dedicated beam time for DVCS measurements with COMPASSII is foreseen for 2016-2017, 
which will provide further constraint on the DVCS cross section in the low-$x$ region.

Depending on the polarization degrees of freedom active in the DVCS process, different GPDs can be 
accessed. 
Some of the GPDs reduce, in the forward limit, to the usual 1-D PDFs. The analysis of the $-t$ 
dependence of 
these GPDs can then shed light on how the related charges are distributed inside the nucleon. 
A recent measurement by the Hall-B (CLAS) Collaboration realized on a longitudinally-polarized ${\rm NH}_3$ 
target~\cite{Seder:2014cdc,Pisano:2015iqa} compared the $-t$ dependence for the beam-spin asymmetry 
(related to the GPD $\mathrm{H}$, that reduces to the electric charge) to the target-spin asymmetry 
(related to the GPD $\mathrm{\tilde{H}}$, that reduces to the axial charge). 
The $-t$ dependence observed for the beam-spin asymmetry is steeper than that for the target-spin 
asymmetry,
thus suggesting that the axial charge is focused inside the nucleon volume more than the electric one.

A complementary 3-D representation of the nucleon is provided, in the momentum space, by 
transverse-momentum-dependent parton distributions (TMD-PDFs).
These are a generalization of PDFs, which include an additional dependence on a small component 
$\mathbold{k_T}$ of the parton momentum, transverse to the virtual photon direction. 
Like PDFs and GPDs, TMDs also depend on the usual longitudinal momentum fraction $x$, and are 
accessed 
through hard processes, in particular in SIDIS, electron-positron annihilation and $pp$ collisions.

In SIDIS, in addition to the outgoing electron, at least one hadron is detected. 
The kinematic of the latter is then used to explore the dynamics characterizing the 
struck quark motion inside the nucleon, by relating the transverse hadron momentum $\mathbold{P_T}$ 
to the (not accessible) parton $\mathbold{k_T}$. In order to describe the mechanism leading to the 
formation 
of the final hadron through the electron scattering off a given target, a second, nonperturbative, 
object is needed in the cross section, the transverse-momentum-dependent fragmentation functions 
(TMD-FFs). 
At leading twist, the SIDIS cross section can be expressed in terms of up to eighteen structure 
functions
(with each structure function being a convolution over $\mathbold{k_T}$ of a TMD-PDF and a 
TMD-FF)~\cite{Anselmino:2011ch}. Each of these structure functions can be accessed through specific 
combinations of polarization degrees of freedom (\textit{e.g.}, unpolarized, 
longitudinally polarized or transversely polarized beam or target). 

\begin{figure}[t]
\centering
\includegraphics[width=\textwidth]{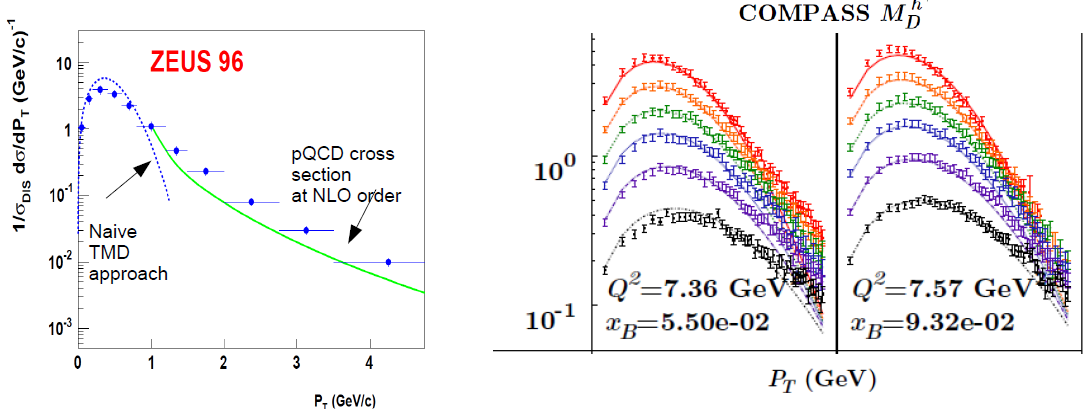}\\
\caption{The dependence of the DIS cross section on the hadron transverse 
momentum $P_T$ for two different kinematics regimes. 
ZEUS data on the left plot show deviation from the Gaussian shape in 
the high-$P_T$ region, while COMPASS multiplicities are well reproduced 
in the righ plots.}
\label{fig:xs_pt}
\end{figure}

The unknown $\mathbold{k_T}$ dependence needed to disentangle the two nonperturbative functions is an 
essential object in
describing the parton dynamics. However, it is not directly accessible, so constraints on it can only 
come through
phenomenological tests on the different hypothesis on its shape.
Furthermore, providing theoretical predictions of cross sections for hadronic processes over the 
whole 
$P_T$ range explored by experiments is a highly nontrivial task. Fixed-order pQCD fails to describe 
low-$P_T$ data, and the cross section tail at large $P_T$ seems to deviate from the usual Gaussian 
ansatz, see Fig. \ref{fig:xs_pt}. A recent review on the transverse spin phenomenology can be found in 
Ref.~\cite{Boglione:2015zyc}.

SIDIS data have been collected during the last decades by several experiments, providing first
constraints on the various TMD-PDFs and TMD-FFs. 
Recently, an extended lever-arm in $Q^2$ has been explored, thanks to simultaneous measurements 
by COMPASS in the low-$x$
region and by HERMES and JLAB in the medium and high-$x$ regime.
Among all the various observables, the transverse target-spin asymmetry $A_{UT}$ plays a crucial role, 
since it provides access to the Collins and Sivers amplitudes.
The former is essential since it is one of the possible ways to access the transversity PDF 
$h_1(x)$: indeed, being chiral-odd, $h_1(x)$ cannot be accessed through inclusive DIS. 
Instead, it has to appear in observables coupled to a second chiral-odd object, which 
is represented by the Collins function in SIDIS.

\begin{figure}[t]
\centering
\includegraphics[scale=0.65]{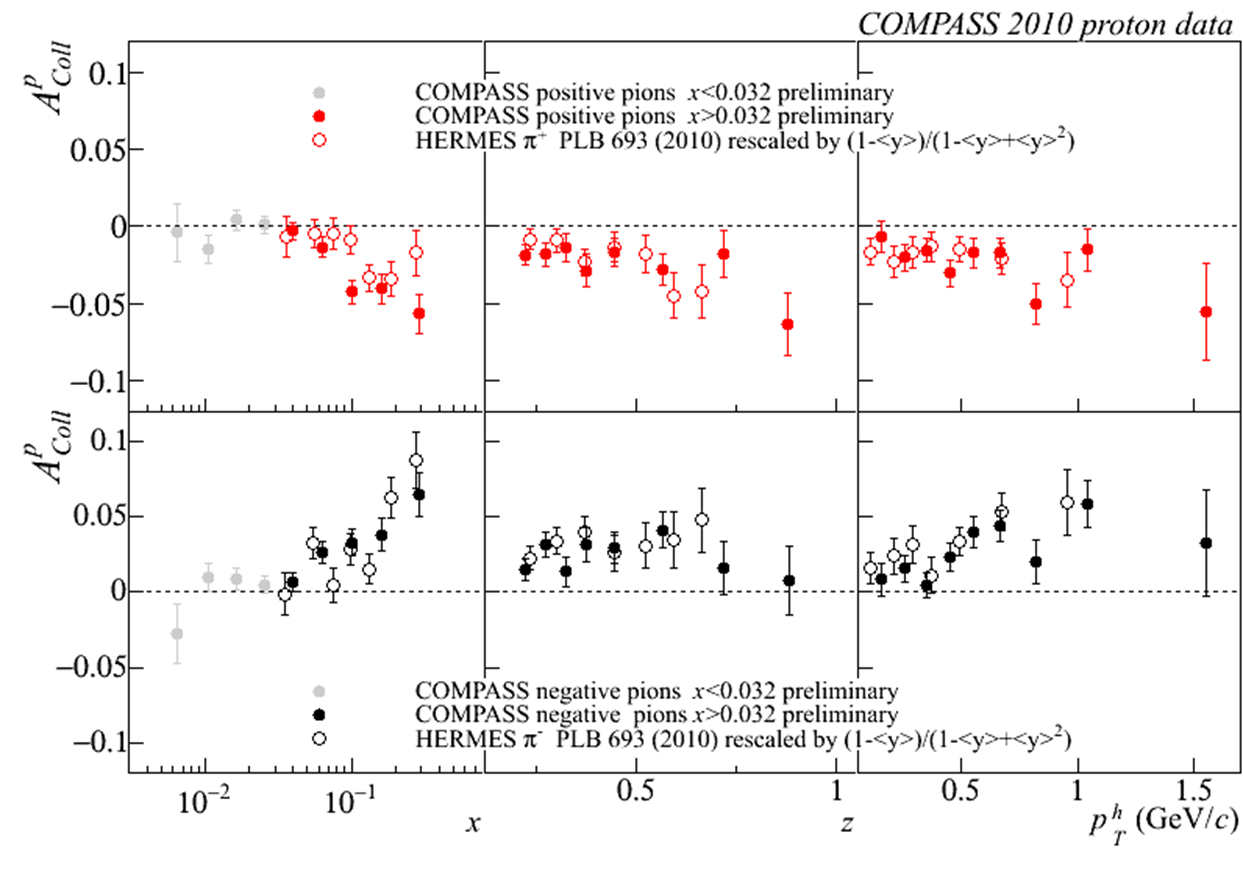}\\
\caption{The Collins modulation of the proton, $A^p_{Coll}$, as a function of $x$, $z$ and $P_T$ for
positive pions (top) and for negative pions (bottom) as extracted from COMPASS and HERMES 
\cite{Airapetian:2010ds}.}
\label{fig:collins}
\end{figure}

A comparison between the Collins modulation on the proton extracted by COMPASS~\cite{Adolph:2014zba} 
and 
HERMES~\cite{Airapetian:2010ds} is shown in Fig.~\ref{fig:collins}. The two sets of data cover 
different kinematic domains, and show a good agreement in the overlapping regions.
By combining proton and deuteron data, a flavor separation can be achieved, possibly leading to 
the extraction of the transversity for the single quark flavors~\cite{Martin:2014wua}.\\

As far as electron-positron annihilation is concerned, first BELLE measurements of double 
differential 
cross sections of two charged pions and kaons as a function of the fractional energies of the two 
hadrons for any
charge and hadron combination have been presented~\cite{Seidl:2015lla}. The ratios of these 
di-hadron cross 
sections for different charges and hadron combinations directly shed light on the contributing 
TMD-FFs. For example, it was found that the ratio of same-sign pion pairs over opposite-sign 
pion pairs drops toward higher fractional energies, where disfavored fragmentation is expected to be 
suppressed. 

\begin{figure}[!t]
\centering
\includegraphics[width=\textwidth]{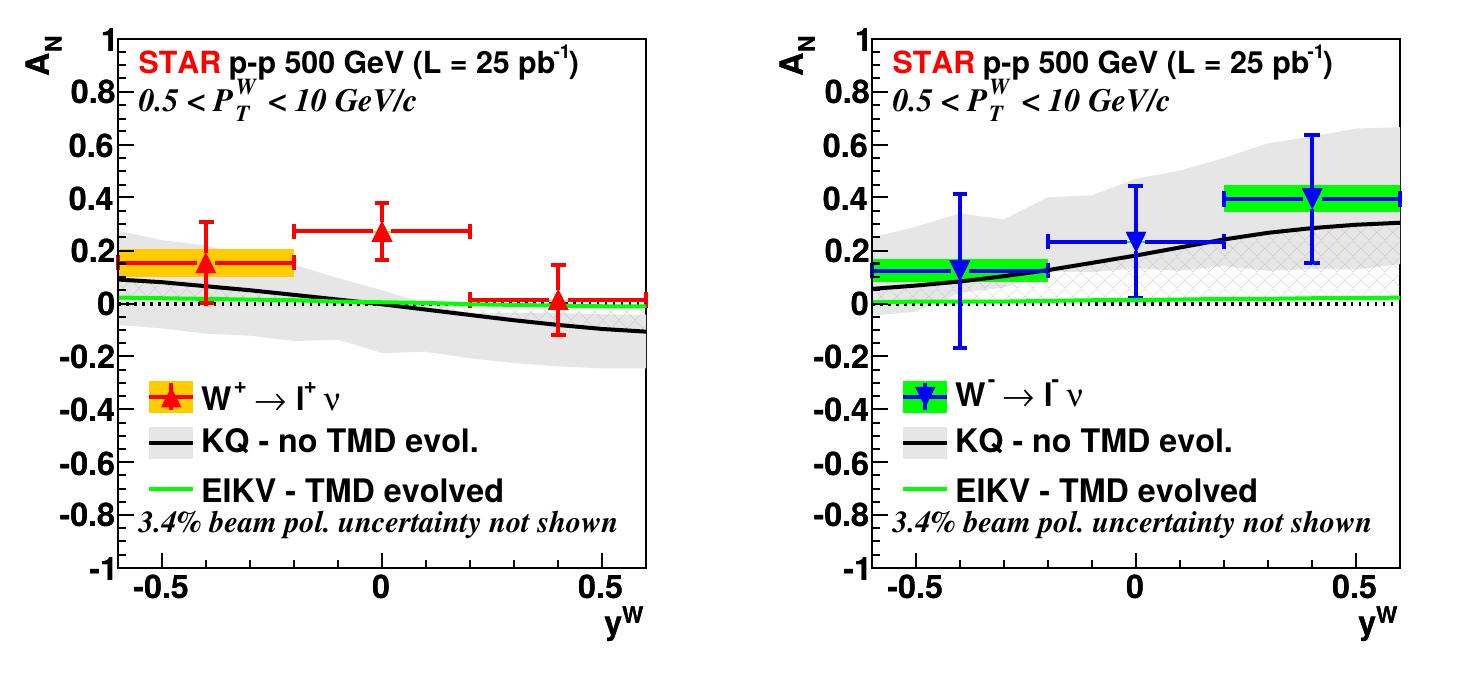}\\
\caption{The transverse single-spin asymmetry $A_N$ for $W^\pm$ and $Z^0$ boson production measured by
STAR in $pp$ collisions at $\sqrt{s}=500$ GeV with a recorded luminosity of 25 pb$^{−1}$. 
The solid gray bands represent the uncertainty on the KQ model~\cite{Kang:2009bp} due to the 
unknown sea quark Sivers function. The crosshatched region indicates the
current uncertainty in the theoretical predictions due to TMD evolution. This figure is taken from
Ref.~\cite{Adamczyk:2015gyk}.}
\label{fig:sfazio}
\end{figure}

\begin{figure}[!t]
\centering
\includegraphics[width=\textwidth]{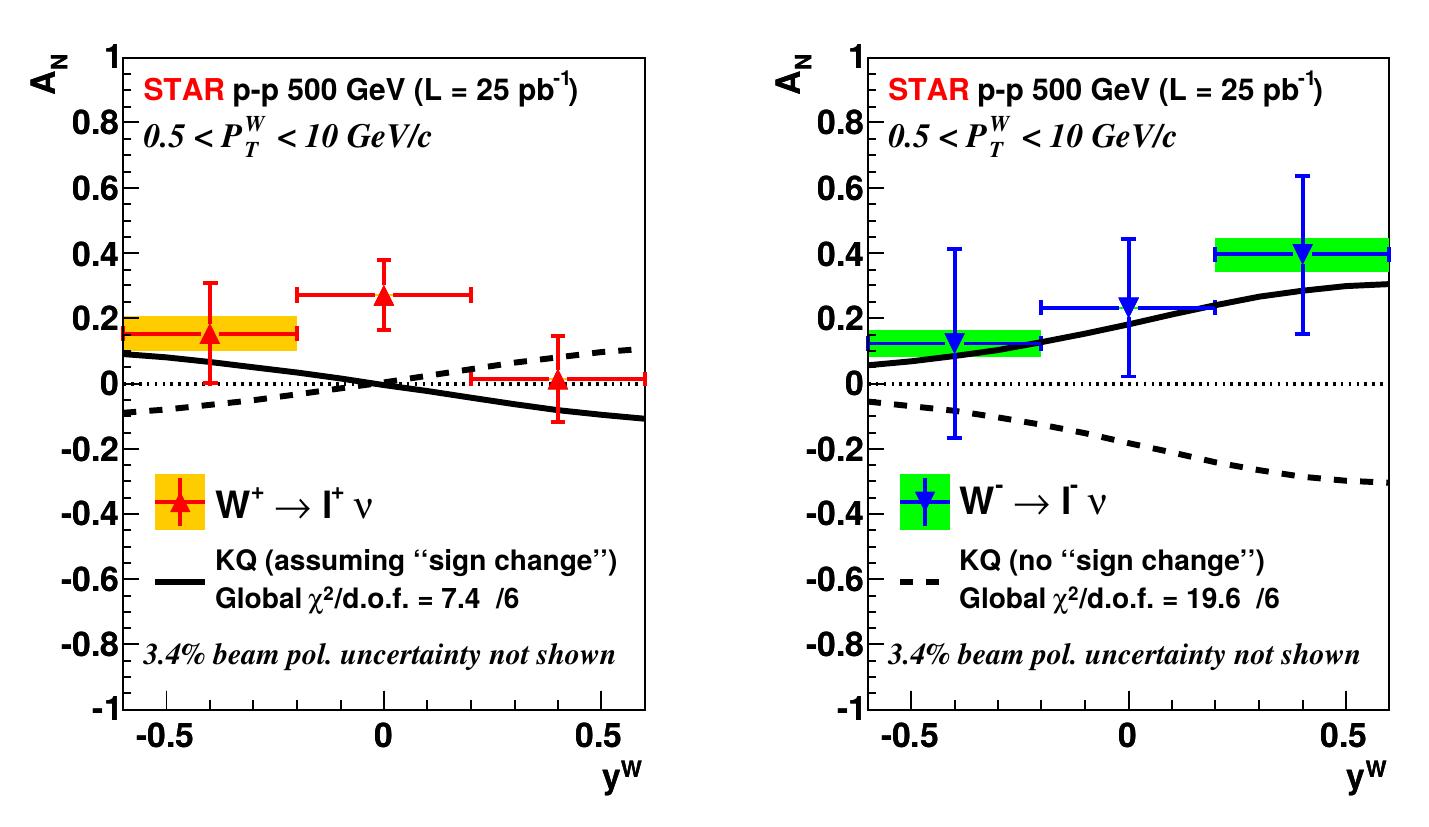}\\
\caption{The transverse single-spin asymmetry $A_N$ for $W^+$ (left plot) and $W^−$ (right plot)
boson production as a function of the boson rapidity $y^W$ compared
to the non TMD-evolved KQ model~\cite{Kang:2009bp}, assuming (solid line) or excluding (dashed line)
a sign change in the Sivers function. Figure taken from Ref.~\cite{Adamczyk:2015gyk}.}
\label{fig:sfazio1}
\end{figure}

Finally, significant progress has also been reported in $pp$ collisions. Most importantly, a 
first measurement of the transverse single-spin asymmetry of Drell-Yan (DY) weak boson production
in transversely polarized $pp$ collisions at $\sqrt{s}=500$ GeV by the STAR experiment has 
been discussed~\cite{Adamczyk:2015gyk}. The measured observable is sensitive to the Sivers function 
$f_{1T}^{\perp}$~\cite{Sivers:1989cc}, one of the TMDs which
describes the correlation between the intrinsic transverse
momentum of a parton and the spin of the parent proton.
A general prediction following from Gauge invariance of QCD and QCD factorization 
formalism~\cite{Collins:2002kn,Mulders:2016dis} is that TMDs in general are not universal objects,
but rather depend on the process under examination. For instance, the Sivers function 
is predicted~\cite{Collins:2002kn} to have the opposite sign in SIDIS compared to 
processes with color changes in the initial state and a colorless final state, such as 
$pp\to DY/W^\pm/Z^0$. Results for the transverse spin asymmetry for weak boson production,
$A_N$ are shown in Fig.~\ref{fig:sfazio} as a function of the boson rapidity $y^W$.
Theoretical predictions based on the KQ~\cite{Kang:2009bp}
and on the EIKV~\cite{Echevarria:2014xaa} models are also displayed. The latter includes TMD
evolution effects, and predicts the largest effects of TMD evolution among many 
TMD-evolved theoretical computations. It was shown~\cite{Adamczyk:2015gyk} 
that a combined fit to $W^+$ and $W^-$ transverse spin asymmetries based on the KQ model
leads to a significantly better $\chi^2$ if a sign-change of the Sivers function is assumed, 
see Fig~\ref{fig:sfazio1}. 
Therefore, provided TMD evolution effects are small, the current 
data seems to favor theoretical models that include a change of sign for the 
Sivers function in comparison to observations in  SIDIS  measurements. 


\section*{Acknowledgements}

We were glad to serve as conveners of the DIS2016 Spin Working Group. 
We would like to warmly thank all the speakers for their interesting talks and discussions, and
the local organizing committee for arranging a fruitful conference and a pleasant stay in Hamburg. 
E.R.N. is supported by a STFC Rutherford Grant ST/M003787/1.

\end{document}